# Real-time spectral analysis of ultrafast pulses using a free-space angular chirp enhanced delay

YiqingXu, [1-3,*] and Stuart G. Murdoch[2, 3]

[1]*College of information science and electronic engineering, Zhejiang University, Hangzhou, Zhejiang, P. R. China*

[2]*The Dodd-Walls Centre for Photonic and Quantum Technologies, Auckland 1010, New Zealand*

[3]*Department of Physics, University of Auckland, Auckland 1010, New Zealand*

*Corresponding author: yxu079@aucklanduni.ac.nz*

**Abstract:** Frequency to time mapping is a powerful technique for observing ultrafast phenomena and non-repetitive events in optics. However, many optical sources operate in wavelength regions, or at power levels, that are not compatible with standard frequency to time mapping implementations. The recently developed free-space angular chirp enhanced delay (FACED) removes many of these limitations, and offers a linear frequency to time mapping in any wavelength region where high-reflectivity mirrors and diffractive optics are available. In this work, we present a detailed formulation of the optical transfer function of a FACED device. Experimentally, we verify the properties of this transfer function, and then present simple guidelines to guarantee the correct operation of a FACED frequency to time measurement. We also experimentally demonstrate the real-time spectral analysis of femtosecond and picosecond pulses using this system.

Frequency to time (F2T) mapping is a simple but powerful technique used to probe the real-time spectral evolution of ultrafast optical systems [1]. At its base, the idea is extremely simple. An ultrafast optical signal is subjected to a large linear group-velocity-dispersion (GVD) [2], this transforms the intensity of the signal's output temporal waveform into a replica of its input optical spectrum, allowing for the direct real-time measurement of its spectral evolution. Variants of this technique have been used to characterize optical rogue waves [3, 4], the onset of mode-locking [5, 6], and other statistically rare optical events that occur on timescales too fast to be captured by conventional electronic digitizers [7-9]. In addition, F2T mapping can be readily adapted to measure other dynamic properties in real-time. For example, real-time optical microscopes [10], and spectroscopy systems [11], based around the same time-stretch principle have also been demonstrated. All these systems require a mechanism to impart a large quadratic spectral phase shift onto the input optical signal. By far the most common way to do this is through the use of a long length of single mode optical fiber, though chirped Bragg gratings [12], arrayed waveguide gratings [13], and more complex 'chromomodal' setups exploiting the large dispersion of multimode optical fibers have also been reported [14].

In this Letter we consider a newly reported device in which the required spectral dispersion is induced by the optical path difference imparted by two tilted high reflectors. This configuration is known as a free-space angular chirp enhanced delay (FACED) [15]. The FACED device consists of a spectrally dispersive element, typically a diffraction grating, followed by two long, high-reflecting mirrors with a slight angle between them. These mirrors impart a large temporal delay across the angularly dispersed input. The advantages of this system over the conventional approaches to F2T mapping are numerous: (1) the system can be operated at any wavelength where dispersive optics and high-reflectivity mirrors are available. This is to be compared to fiber F2T mapping setups which can only supply the large dispersion required at wavelengths where the fiber loss is low, (2) the induced GVD is purely linear with no distortions due to higher-order dispersion, (3) the system is perfectly linear with no distortion, or input power limitations, due to nonlinearities such as the Kerr nonlinearity present in optical fibers, and (4) the amount of dispersion imparted to the input signal is set by the distance between the FACED mirror pair and can be made very large. In this work we demonstrate the successful measurement of the real-time spectra of both femtosecond and picosecond optical pulse sources operating around 1550 nm using a FACED device. We note that, the dispersion the FACED system imparts to measure the 4 ps pulses used in our second experiment is 3ns/nm. Achieving a similar dispersion in a fiber F2T system would require the use of 150 km of standard optical fiber.

The primary complication of the FACED device is that the dispersed output signal is not spectrally continuous but rather sampled at a frequency spacing set by the cardinal modes of the mirror pair. As a consequence, it is necessary to consider in detail the spectral and temporal sampling characteristics of the FACED device. In this Letter we present the first detailed analysis of the spectral transmission of this system, and present simple guidelines to enable its correct operation. We organize this Letter as follows: we first introduce the basic configuration and the key parameters of the FACED setup, and outline how the F2T mapping process is related to these parameters. We derive the spectral transfer function of the FACED, and experimentally confirm these characteristics using an ASE source. We then present F2T measurements of both femtosecond and picosecond pulses, and investigate the effect of the angle and spacing of the FACED mirrors on these measurements.

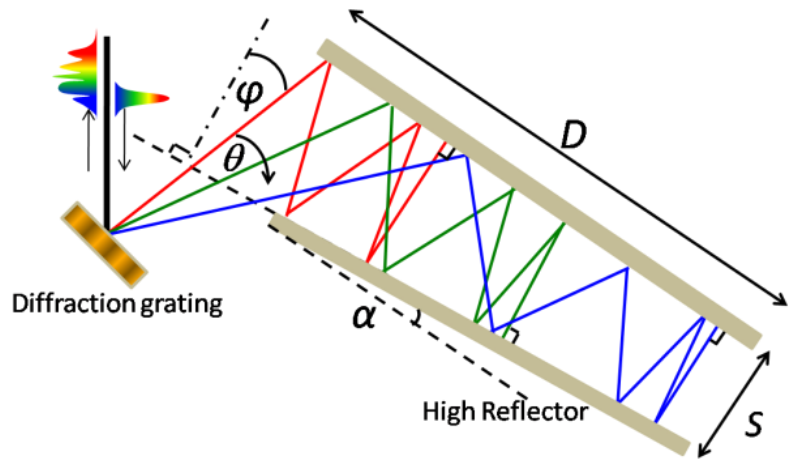

**Fig.1**. A simplified schematic of a FACED cavity. The ray diagram illustrates the different possible optical paths of an angularly dispersed optical beam propagating inside the mirror pair.

We consider the FACED geometry first proposed in Ref. [15], and simplify this scheme by removing the 4-f beam relay system. A schematic diagram of our setup is shown in Fig. 1. The input beam is first incident on a diffraction grating which angularly disperses the input spectrum. This is followed by two high reflectors of length $D$, spaced by a separation $S$, and tilted with respect to each other by an angle $\alpha$. As shown in Fig. 1, only a discrete set of wavelengths whose input angles match the cardinal modes of the FACED mirrors will be reflected back and captured by the numerical aperture of the collection optics. Three such representative wavelengths are shown in Fig. 1 as red, green and blue rays. All other wavelengths are not captured by the photodetector at the receiver. This illustrates the key spectral sampling behavior of the FACED system. The tilt angle between the two high reflectors sets the rate at which the input spectrum is sampled, with adjacent cardinal modes spaced in angle by $\alpha$. Defining $\Delta\theta$ as the total angular range of the dispersed spectrum exiting the diffraction grating thus sets the total number of cardinal modes (or sampled wavelengths) as $N=\Delta\theta/\alpha$. For a small tilt angle, the group delay between adjacent cardinal modes can be well approximated as $Tg= 2S/c$, where $c$ is the speed of the light. Thus, the total dispersed temporal width of the output signal is simply equal to $\tau_{width} = NT_g$. Accordingly, the total dispersion induced by the FACED mirrors can be written as:

$$D_{total} = NT_g / \Delta\lambda = (2S/c\alpha)(\Delta\theta/\Delta\lambda) \tag{1}$$

where $\Delta\lambda$ is the spectral width of the input signal. At this point it is important to note that this dispersion parameter plays a different role in a FACED device than in a standard F2T setup. Ina conventionalF2T mapping measurement, the total amount of induced dispersion directly sets the spectral resolution of the measurement. However, the spectral resolution of a correctly optimized FACED F2T measurement depends only on the spectral resolution of the dispersive element used. Two separate conditions are required to ensure the correct operation of a FACED F2T measurement. The first requirement is that the temporal separation between adjacent cardinal modes must be larger than the temporal resolution of the digitizer used, such that we can temporally resolve each cardinal mode (i.e., $Tg> 1/B$, where $B$ is the electronic bandwidth of the digitizer used).To achieve this first requirement, the spacing $S$ between the two mirrors must be set sufficiently large. Secondly, the angularly dispersed spectrum of the input pulse needs to be correctly sampled. As the spectral resolution of a FACED device is set by the wavelength resolution $\delta\lambda$ of the dispersive element, the spectral sampling rate $fs = \alpha(\Delta\lambda/\Delta\theta)$ of the FACED mirrors should be set to be of the order of $\delta\lambda$ or smaller. The minimum number of cardinal modes required to ensure accurate spectral sampling can thus be written as:

$$N_{min} = \frac{\Delta\theta}{\alpha} \geq \frac{\Delta\lambda}{\delta\lambda} \tag{2}$$

This second requirement can always be achieved by setting the angle between the two mirrors $\alpha$ to be sufficiently small. Thus, for a correctly aligned FACED system, the spectral resolution $\delta\lambda$ is determined only by the diffractive element, and the spacing S, and tilt angle α, of the mirror pair must be set according to the two preceding requirements. This result also emphasizes another difference between the FACED system and conventional F2T measurements. In a standard measurement the signal power of any dispersed spectral element is directly proportional to the inverse of the applied dispersion. Thus, an F2T system with a large amount of dispersion, will necessarily have proportionally lower power in each element of the output temporal signal. In a FACED system, however, the signal power in the temporally dispersed signal scales directly as *1/N*. Increasing the dispersion of the device by increasing the spacing S between the mirror pair will have no effect on the signal power in each spectral sample.

Finally, before we proceed to the experimental verification of these results, we present the spectral transfer function of the FACED system. The transfer function in the spectral domain can be simply expressed as the summation of the reflection of all cardinal modes.

$$R(f) = \sum_{n=1}^{N} \{\delta[f - f(n\alpha)] \otimes Tr(f', \delta\lambda) \cdot \exp(j\Phi(f))\} \tag{3}$$

where the frequency *f(θ)* is a function of the beam angle θ which can be calculated from the diffraction grating equation. The spectrum is sampled at integer multiples of the tilting angle, with *r(f,δλ)* the reflection profile of a single cardinal mode with a spectral width *δλ*. The final term *Φ(f)* accounts for the phase shift of each of the cardinal modes induced by the change in optical path length:

$$\Phi(f) = \frac{2\pi f}{c} OPL(f) \tag{4}$$

where the optical path length *OPL(f)* is the chord length in the conjugated mirror model introduced in Ref. [15],

$$OPL[\theta(f)] = \frac{2S}{\alpha} \sin\frac{\theta(f)}{2} \tag{5}$$

If required, the temporally dispersed optical field exiting the FACED device can then be calculated from this transfer function by multiplying Eqn. (4) by the spectral field of the input signal and taking the inverse Fourier transform of the product.

In the experimental results presented below we verify the FACED spectral transfer function and perform a characterization of FACED F2T mapping using both femtosecond and picosecond pulses. We note that, whilst we chose to perform our measurements at 1550 nm, this is simply for convenience. One of the primary advantages of the FACED system is its ability to operate in different spectral regions (such as the visible) where low-loss optical fibers are not available. We first characterize the transfer function of the FACED system using an amplified spontaneous emission (ASE) source derived from an Erbium doped fiber amplifier (EDFA).The ASE exiting the fiber is collimated using a 4.5 mm focal length aspheric lens (Thorlabs C230TMD-C), before passing through a 50/50 beam splitter. The collimated beam is dispersed by a diffraction grating with a groove density of 900 line/mm, and this spectrally dispersed signal is then launched into the FACED mirror pair. This consists of two 15cm long high reflectors (with a reflectivity of ~99.5% around 1550nm). The spacing of the mirror pair is initially set to S = 30 mm, and the tilt angle to α = 0.7 mRad. After traversing the FACED mirrors, frequencies corresponding to cardinal modes of the system are retro-reflected and exit via the reflection port of the input beam splitter. This returned beam is coupled back into a multimode fiber (core size of

62.5 µm) via another 4.5 mm focal length aspheric lens. The total loss of the system at an input frequency corresponding to a cardinal mode of the system is 10.5 dB. This loss comprises of, a 6 dB loss from passing through the beam splitter twice, a 3 dB loss from a double pass of the diffraction grating, and 1.5dB fiber coupling loss into the multimode fiber. We note that, using more complex out-coupling optics (e.g. a vertically offset output) the 6 dB loss of the beam splitter could be eliminated resulting a very low total loss for anF2T system of only 4.5 dB. The output spectrum from the FACED is measured using an optical spectral analyzer (OSA, Yokogawa AQ6370).In Figs 2. (a–c) respectively, we plot the measured experimental spectral profiles reflected from the FACED system at three different tilt angles, $\alpha$ = 0.7, 1, and 1.6 mRad. The corresponding theoretical spectral transfer functions calculated using Eq. (4) are shown in Figs. 2 (d–f).In these theoretical curves, we have set the reflection profile of a single cardinal mode to be Gaussian with a spectral width (resolution) of $\delta\lambda$=0.4 nm. The agreement between these two sets of curves is excellent. As can be seen from Fig. 2, the spacing between adjacent cardinal modes is easily controlled by adjusting the mirrors tilt angle, with smaller angles yielding a smaller spacing between the modes. By this means it is always possible to set the mode spacing of the FACED mirrors to ensure the spectral features of the input optical signal will not be under sampled.

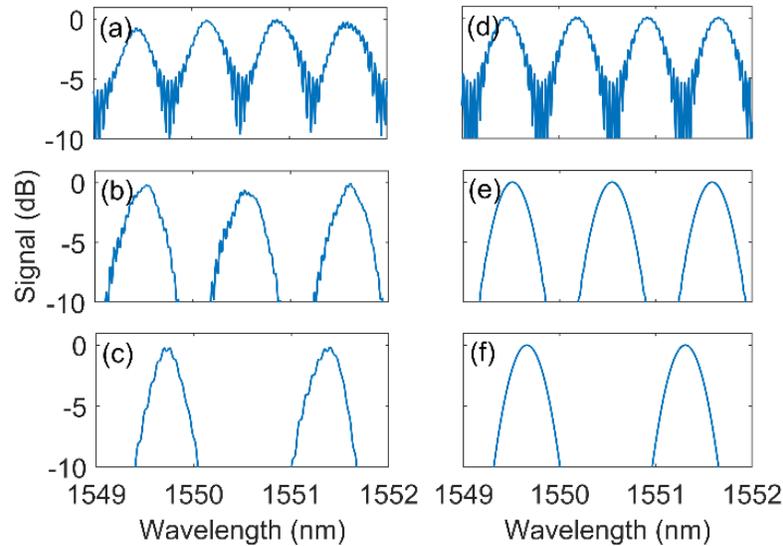

**Fig. 2.** (a–c) Experimentally measured spectral transfer function of the FACED system at tilt angles $\alpha$ = 0.7, 1, and 1.6 mRad, and (d – f) the corresponding theoretical transfer functions given by Eqn. (4).

We next consider the spectral measurement of a femtosecond mode-locked laser using a FACED F2T mapping system. The ASE source is replaced by a280 fs mode-locked fiber laser centered at 1554 nm with an average power of 2 mW. Due to self-phase modulation in the output fiber patch cord, and the pigtailed fiber isolator which follows the laser, the FWHM of the spectrum of the output pulse is broadened to 14nm and has a double-peaked structure [16]. The temporal intensity signal exiting the FACED is measured using a fast photodetector (12 GHz bandwidth) and real-time oscilloscope (12 GHz bandwidth, 40 GS/s sampling rate). The spectrum of the laser is also measured using a conventional OSA to provide a reference against which to compare the FACED measurement. We measure the output of the FACED device at three different mirror spacings ($S$ = 15, 27.5 and 37.5 mm). The tilt angle is set to ~ 0.2 mRad which sets the number of cardinal modes across the laser's 14 nm bandwidth to N ~ 56. The results of these measurements are plotted in Fig. 3. As can be seen, the output temporal FACED traces match well with the input spectrum (appropriately scaled versions of which are plotted in Fig. 3 as black lines). We note that, the FACED signal of Fig. 3 corresponds to a direct measurement of the spectrum of a single pulse of the laser, with no averaging applied. Fig. 3 clearly demonstrates that increasing the mirror separation S, increases the temporal delay between adjacent cardinal modes, but has no effect the spectral sampling of the input signal with approximately 56 modes visible in each of the three traces shown. Measurements of spectrum of the light exiting the FACED system also clearly demonstrate this, with no significant change in the output spectrum observed as S is varied. This allows the FACED mirror spacing to be set to ensure that the cardinal modes can be individually resolved by the electronic detection system used. The only limit set on the total amount of dispersion possible is the finite length of the mirror pair used.

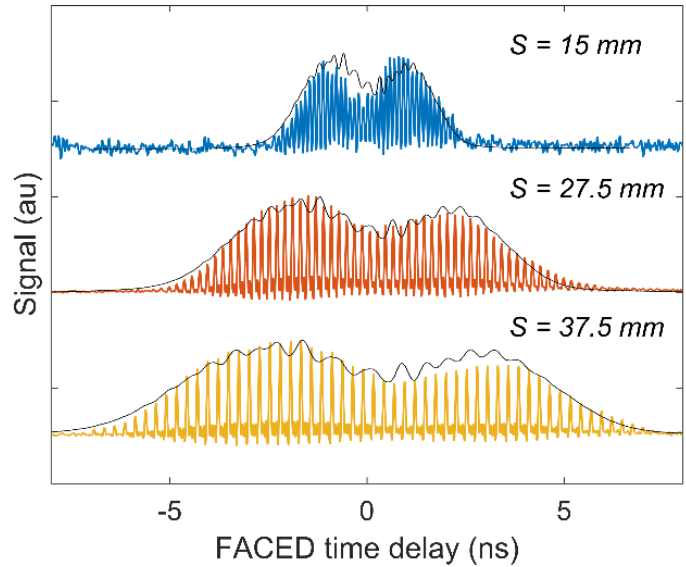

**Fig.3.** The FACED temporal intensity profiles of a 280 fs pulse after F2T mapping, at three different mirror spacings, and a constant tilting angle ($\alpha$ = 0.2 mRad). The fine black curves show the measured spectrum of the input signal appropriately scaled to match the FACED curves.

To further demonstrate the capability of FACED for high-resolution real-time spectral analysis, we next present a characterization of the spectrum of a picosecond pulse train. We obtain the picosecond signal by filtering the previous mode-locked laser with a 100 GHz optical band-pass filter centered at 1554 nm. The filtered pulses are then re-amplified with an EDFA to 50 mW. As a result of this amplification the pulse develops a distinct asymmetric three-lobed spectrum that we wish to characterize using our FACED system. The spectral full-width-half-maximum of this output is 1.5 nm, and the estimated pulse width is 4 ps. To increase the spectral resolution of the FACED device, the input collimation lens is changed to a microscope objective (Melles Griot 4/0.12, f = 16 mm), and the diffraction grating is replaced with a one of 1200 line/mm groove density (Thorlabs GR25-1210), resulting in an improved spectral resolution of the dispersive section of the FACED to $\delta\lambda = 0.05$ nm. In Fig. 4 we plot the temporal intensity profiles of the ps pulse after F2T mapping at three different tilt angles between the two FACED mirrors. This figure clearly illustrates the importance of correctly selecting the tilt angle to ensure sufficient spectral sampling of the input spectrum. In the top curve the tilt angle is set to 0.44 mRad, setting the spectral sampling interval to 0.32 nm, and allowing for only 8 cardinal modes within the bandwidth of the pulse. As a result, the three-peaked spectrum of the input signal (shown as the superimposed black trace) is not fully resolved. In the middle and lower traces the tilt angle is lowered to 0.22 and 0.11 mRad respectively. This increases the number of cardinal modes that lie within the spectrum to 16 and 32 and allows a full accurate measurement of the input signal's optical spectrum. Fig. 4 also clearly shows that as the tilt angle is varied, the spacing between adjacent cardinal modes is unchanged. Again, we note that, the FACED traces shown in Fig. 4 correspond to the individual spectra of single pulses of the picosecond source. The dispersion imparted to the optical signal by the FACED system at the lowest tilt angle ($\alpha = 0.11$ mRad) is ~ 3 ns/nm. To achieve a similar dispersion in a conventional fiber F2T setup would require ~ 150 km of standard single mode fiber at 1550 nm, imparting an unsustainable 30 dB loss during the F2T conversion.

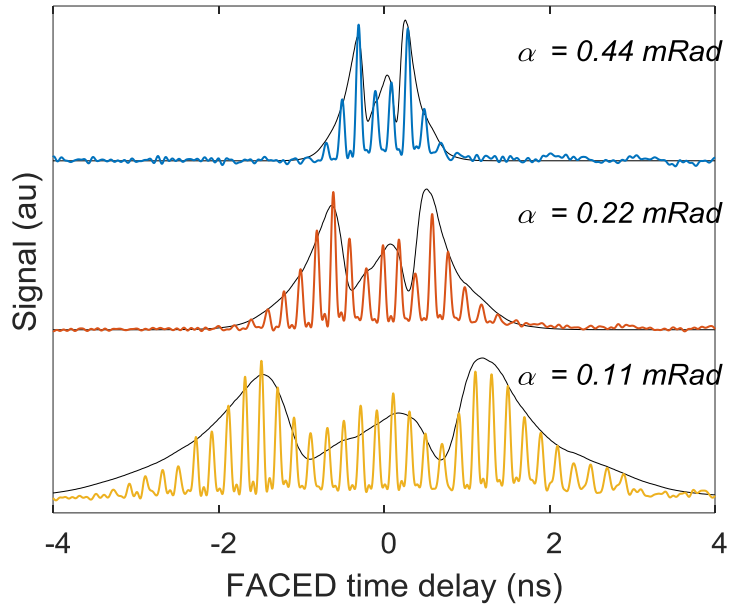

**Fig. 4.** The FACED temporal intensity profile of a 4 ps pulse after F2T mapping, at three different tilt angles, with a spacing S = 30 mm between the mirrors. The fine black curves show the measured spectrum of the input signal appropriately scaled to match the FACED curves.

In summary, we have presented a full analysis of the spectral and temporal properties of a FACED F2T mapping system. We have experimentally verified the spectral transfer function of our FACED device using an ASE source, and demonstrated F2T measurements of the spectra of both femtosecond and picosecond pulse sources. This has allowed us to demonstrate the effect of the key FACED parameters on the output temporal waveforms. The discrete spectral sampling of a FACED measurement is significantly different than many other F2T methods. We provide simple guidelines to ensure the FACED mirror pair is correctly aligned to allow for accurate spectral measurement. The FACED geometry presented in this Letter can be implemented at any wavelength where high-reflectivity mirrors and dispersive elements are available. In addition, its free-space nature means that it is inherently free from many of the distortions that can occur in other F2T implementations. For these reasons, we believe that this new geometry will prove useful for the many applications where standard F2T mapping is not currently practical.

**Funding.** The Dodd-Walls Centre for Photonic and Quantum Technologies, New Ideas Research Fund.